\documentclass{article}

\usepackage[final, nonatbib]{mlncp_2023}
\usepackage[square,sort,comma,numbers]{natbib}
\usepackage[utf8]{inputenc} %
\usepackage[T1]{fontenc}    %
\usepackage{hyperref}       %
\usepackage{url}            %
\usepackage{booktabs}       %
\usepackage{amsfonts}       %
\usepackage{nicefrac}       %
\usepackage{microtype}      %
\usepackage{xcolor}         %
\usepackage{graphicx}
\usepackage{amsmath}
\usepackage{amssymb}
\usepackage{colortbl}
\usepackage{siunitx}
\usepackage{comment}
\usepackage{pifont}
\usepackage{subfig}

\definecolor{a}{rgb}{0.83, 0.83, 0.83}
\definecolor{b}{rgb}{0.99,0.99,0.99}

\title{The Data Conversion Bottleneck in Analog Computing Accelerators}

\author{%
James T. Meech$^{1}$ \quad Vasileios Tsoutsouras$^{1,2}$ \quad Phillip Stanley-Marbell$^{1,2}$ \\
$^1$Department of Engineering, University of Cambridge \quad $^2$Signaloid\\
\texttt{jtm45@cam.ac.uk, vt298@eng.cam.ac.uk, phillip.stanley-marbell@eng.cam.ac.uk }\\
}

\begin{document}

\maketitle

\begin{abstract}
  Most modern computing tasks have digital electronic input and output data.
  Due to these constraints imposed by real-world use cases of computer systems, any analog computing accelerator, whether analog electronic or optical, must perform an analog-to-digital conversion on its input data and a subsequent digital-to-analog conversion on its output data. 
  The energy and latency costs incurred by data conversion place performance limits on analog computing accelerators.
  To avoid this overhead, analog hardware must replace the full functionality of traditional digital electronic computer hardware. 
  This is not currently possible for optical computing accelerators due to limitations in gain, input-output isolation, and information storage in optical hardware.
  This article presents a case study that profiles 27 benchmarks for an analog optical Fourier transform and convolution accelerator which we designed and built. 
  The case study shows that an ideal optical Fourier transform and convolution accelerator can produce an average speedup of $9.4 \times$ and a median speedup of $1.9 \times$ for the set of benchmarks. 
  The optical Fourier transform and convolution accelerator only produces significant speedup for pure Fourier transform ($45.3 \times$) and convolution ($159.4 \times$) applications.
\end{abstract}

\section{Introduction}\label{section:introduction}

Most modern computing tasks are constrained to having digital electronic input and output data.
Mass-produced digital electronic memory is the only off-the-shelf option for data storage.
This constrains the input data to be digital electronic signals. 
Plotting and data visualization software is only widely available for programming languages designed to run on off-the-shelf digital electronic hardware.
The traditional digital electronic computer architecture is better suited to most applications than current application-specific analog computing accelerators.
Directly substituting analog computer architectures for digital computer architectures would therefore be unproductive: 
For the time being, analog computing accelerators must efficiently compute partial or full results for applications dominated by the type of computing operations the accelerators are designed to accelerate.

Any analog computing accelerator operating on digital input data to produce digital output data must perform a digital-to-analog conversion on its input data and a subsequent analog-to-digital conversion on its output data because of the input and output constraints imposed by modern computer systems. 
The only alternative would be to develop an entire software stack to allow the analog hardware to perform all the functions of the traditional digital electronic computer hardware. 
This is not currently possible for optical computing accelerators due to limitations in gain, input-output isolation, and memory.
Modern digital electronic computers spend 62.7\,\% of their energy moving data~\cite{boroumand2018google}. 
Adding computing accelerators that cannot accelerate the entire application exacerbates this existing data movement bottleneck~\cite{boroumand2018google}.
Power delivery requirements trends are placing even more constraints on available pins and memory bandwidth, making the problem worse still~\cite{5993603}.

Figure~\ref{prototype} in Appendix~\ref{appendix:physicalImplementation} shows a prototype analog optical accelerator we designed and built while studying the data movement and data conversion bottlenecks.
Appendix~\ref{appendix:study} contains an optical Fourier transform and convolution computing accelerator case study which shows that the best possible speedup for optical Fourier transform and convolution accelerators is orders of magnitude smaller than that of other popular accelerator architectures. 
This limited upside on possible speedup will continue to be the case even after research advances overcome the data movement bottleneck. 

\section{The Cost of Digitally Interfacing With Analog Computing Accelerators}\label{section:bottlenecks}

\begin{figure}
  \centering
  \includegraphics[width=\textwidth]{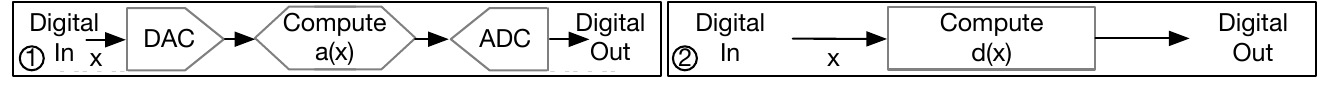}
  \caption{Architectures for computational problems with digital input and output data.
           Let $a(x)$ be an analog function and $d(x)$ be a digital function, both computed using the input data $x$.}\label{fig:fourTypes}
\end{figure}

Most modern computing systems are constrained to having digital electronic input and output data. 
These constraints are imposed by the storage of input data in off-the-shelf digital electronic memory and the data processing and visualization software tools which are exclusively designed to run on digital electronic computer systems. 
To avoid data conversions, computer systems can only use digital computing devices for applications that have digital input and output data. 
Figure~\ref{fig:fourTypes}~block~\ding{192} shows that a computer system must incur the latency and energy cost of a digital-to-analog and analog-to-digital conversion to use analog computing devices for problems with digital input and output data.
Figure~\ref{fig:fourTypes}~block~\ding{193} shows that a computer system does not need any data conversion to perform the same computation using digital computing devices.
Using analog computing device-based accelerators is therefore only worthwhile when the energy and latency saved by using analog computing devices far outweigh the energy and latency costs of the data conversion. 

\begin{figure*}[t]
  \centering
  \subfloat[The power consumption and speed tradeoff for 96 different digital to analog converter designs published in various venues since 1996~\cite{dac_survey}.\label{fig:DACparetoFrontierISSCC}]{%
        \includegraphics[width=0.49\textwidth,keepaspectratio,trim={2cm 7cm 10.5cm 13cm},clip]{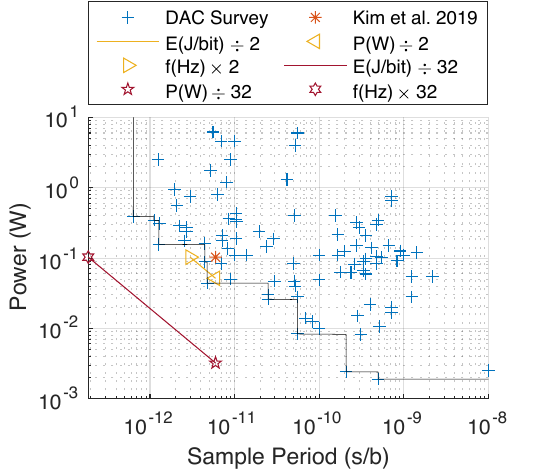}}\hfill
  \subfloat[The power consumption and speed tradeoff for 647 different analog to digital converter designs published in the ISSCC and VLSI conferences since 1996~\cite{adc_survey}.\label{fig:ADCparetoFrontierISSCC}]{%
        \includegraphics[width=0.49\textwidth,keepaspectratio,trim={2cm 7cm 10.5cm 13cm},clip]{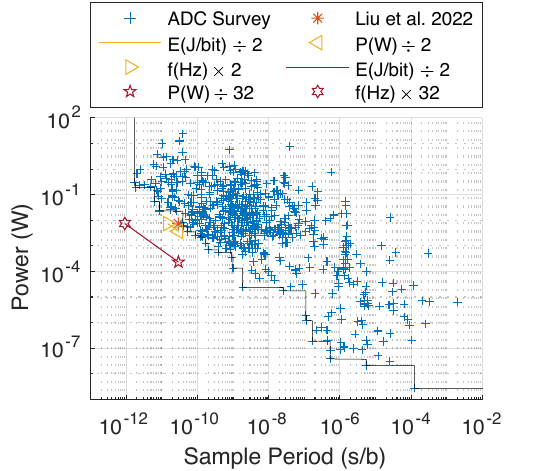}} \\
  \caption{The digital-to-analog and analog-to-digital converter speed and power consumption tradeoff.}\label{prototype}
\end{figure*}

Figure~\ref{fig:DACparetoFrontierISSCC} shows a plot of the sampling speed and power consumption of 96 digital-to-analog converter designs published in various venues including the International Solid-State Circuits Conference (ISSCC) and Symposium on Very Large Scale Integration (VLSI) Technology and Circuits conference between 1996 and 2021. 
Figure~\ref{fig:ADCparetoFrontierISSCC} shows a plot of the sampling speed and power consumption of 647 analog-to-digital converter designs published in ISSCC and VLSI since 1997. 
The Pareto frontier (black stepped line) shows that there is a tradeoff between power consumption and sampling speed for both digital-to-analog and analog-to-digital converters~\cite{dac_survey,adc_survey}. 

Anderson et al.~\cite{anderson2023optical} use values from existing digital-to-analog (Kim et al.~\cite{kim2019four}) and analog-to-digital (Liu et al.~\cite{liu2022adc}) converters which are above the Pareto frontiers of Figure~\ref{fig:DACparetoFrontierISSCC}~and~\ref{fig:ADCparetoFrontierISSCC} to predict an energy efficiency improvement of $100 \times$ over digital electronic hardware for an optical computing accelerator which uses existing technology.
Anderson et al.~\cite{anderson2023optical} predict a greater than $100,000 \times$ energy advantage when performing multiply-accumulate (MAC) operations for an analog optical multiply-accumulate computing accelerator over existing 300fJ/MAC digital electronic hardware (NVIDIA A100 GPU).
The greater than $100,000 \times$ predicted energy advantage relies on the availability of analog-to-digital and digital-to-analog converters which use $32 \times$ fewer joules per bit than Kim et al.~\cite{kim2019four} and Liu et al.~\cite{liu2022adc}, respectively.
Using fewer bits of precision to reduce analog-to-digital and digital-to-analog converter requirements is promising, but any precision reduction tradeoff to reduce power consumption and increase optical hardware speed can be made more easily with digital hardware due to the mature, low-cost manufacturing processes~\cite{9256623, edwards2020google}.
Computer architects should therefore avoid converting a given signal from digital to analog and vice versa unless it is completely necessary.

Figure~\ref{fig:DACparetoFrontierISSCC} shows that reaching the $32 \times$ smaller digital-to-analog converter energy by reducing converter power consumption, increasing sampling speed, or some combination of the two requires a design significantly below and in some cases more than an order of magnitude below the Pareto frontier. 
Figure~\ref{fig:ADCparetoFrontierISSCC} shows that reaching the $32 \times$ smaller analog-to-digital converter energy by reducing converter power consumption, increasing sampling speed, or some combination of the two requires a design more than an order of magnitude below the Pareto frontier. 
Halving the analog-to-digital converter energy target requires moving to a design space that is entirely below the Pareto frontier. 
Implementing a design more than an order of magnitude below the Pareto frontier may not be possible. 

Reducing digital to analog converter power consumption and increasing sampling speed are well-researched topics~\cite{jang2023design,dac_survey,adc_survey}.
Researchers designing and building novel analog computing accelerators should collaborate with digital-to-analog and analog-to-digital converter designers to determine the feasibility of reaching this design point with existing technology.
Fundamentally new methods and devices could be required to meet this goal.
Jang et al.~\cite{jang2023design} state that the best-reported analog-to-digital converter efficiency has improved by nearly six orders of magnitude over the past 40 years.
Jang et al.~\cite{jang2023design} however also state that energy-efficient analog-to-digital converters have low bandwidth.
This is problematic for analog computing accelerator designers as they require high bandwidth analog-to-digital converters to avoid the data movement bottleneck.
Analog computing accelerator designers should collaborate with the ISSCC and VLSI communities to produce faster and more efficient digital-to-analog and analog-to-digital converter designs and implementations.

A research implementation of an optical computing accelerator that mitigates the analog-to-digital and digital-to-analog conversion bottleneck exists~\cite{kalinin2023analog}.
The implementation minimizes the data conversions required by iteratively solving entire optimization problems in the analog domain without repeated digital-to-analog and analog-to-digital conversions. 
The architecture only converts the input data from digital to analog once at the start of the problem and from analog to digital once at the end of the problem.
Their approach has the weakness described in Section~\ref{section:keyes} that the accelerator has to replace more of the functionality that is traditionally implemented using digital electronic hardware.
This leads to limits on the problem sizes that their implementation can solve imposed by the off-the-shelf optical hardware they used. 
The speed at which circuits can operate is determined by their resistance $R$, capacitance $C$, and inductance $L$.
As the implementation needs to convert optical analog signals to electronic analog signals there will be a new bottleneck imposed by the $RC$ and $LC$ delays in the conversion circuitry.
These conversion costs will be lower than analog-to-digital and digital-to-analog conversion costs but we leave a detailed analysis as future work.

\section{Digital Hardware is Required to Facilitate Analog Computing Accelerators}\label{section:keyes}

For the example of optical transformers the data conversion bottleneck is exacerbated because the technology to efficiently implement non-linear activation functions optically does not exist~\cite{anderson2023optical, kalinin2023analog}.
This makes it necessary to transduce the optical signal to an electronic signal, perform an analog to digital conversion, compute the activation function on digital electronic hardware, perform a digital to analog conversion, and finally an electronic to optical signal conversion~\cite{anderson2023optical}.
Performing these conversions for every layer of a neural network makes the resulting computer architecture slow and inefficient, only producing energy savings for greater than ten billion parameter models with current conversion technology~\cite{anderson2023optical, adc_survey, dac_survey}.
Keyes~\cite{keyes1985what} stated in 1985 and Tucker~\cite{tucker2010role} stated again in 2010 that a good computer device requires:

\begin{enumerate}
\item \textbf{Gain:} The ability to produce an output signal larger than the input signal.
\item \textbf{Input-output isolation:} The output of the computing device does not affect the input.
\item \textbf{Information storage:} An efficient and reliable memory cell design.
\end{enumerate}

At the time of writing, digital electronic transistors are the only mass-produced computing devices satisfying these criteria. 
Therefore, analog computing accelerators need digital hardware to interface with the digital computer system providing the input data, postprocessing, and storing the output data.

\subsection{Case Study: Analog Optical Fourier Transform and Computing Accelerator}

It is unclear whether or not existing optical computing hardware can efficiently implement gain. 
Currently, optoelectronic gain (using electronic transistors) is the most common choice in optical neural network research prototypes~\cite{shastri2021photonics,mcmahon2023physics}.
It is unclear whether or not existing optical computing hardware can implement input-output isolation~\cite{anderson2023optical},
but some ongoing efforts attempt to address this challenge by solving an optimization problem that quantifies light leakage~\cite{akhetonics}.
Spatial light modulators do not have input-output isolation. 
Anderson et al.~\cite{anderson2023optical} report that spatial light modulator pixels have crosstalk if their neighbors have significantly different values. 
Anderson et al.~\cite{anderson2023optical} remedied this crosstalk by aggregating $3 \times 3$ blocks of pixels together as macro pixels.
This aggregation is undesirable as it reduces the total number of pixels available for computation by a factor of nine.
Research implementations of integrated optical static random access memory exist with footprints close to those of electronic memories, lower access times, and total energy costs per bit~\cite{alexoudi2020optical,mcmahon2023physics}.
These research implementations of a few bytes must be scaled and integrated into memory architectures as cells that work reliably despite thermal instability and crosstalk.
Optical computer systems are therefore application-specific computing accelerators until they can optically implement gain, input-output isolation, and information storage at scale.
Optical computing hardware currently cannot replace the entire functionality of the digital electronic processor and therefore will only offload selectively-chosen parts of application programs to the optical computing accelerator.

\section{Accelerators Should Target High Complexity Computational Problems}

\begin{figure}
  \centering
  \includegraphics[width=\textwidth]{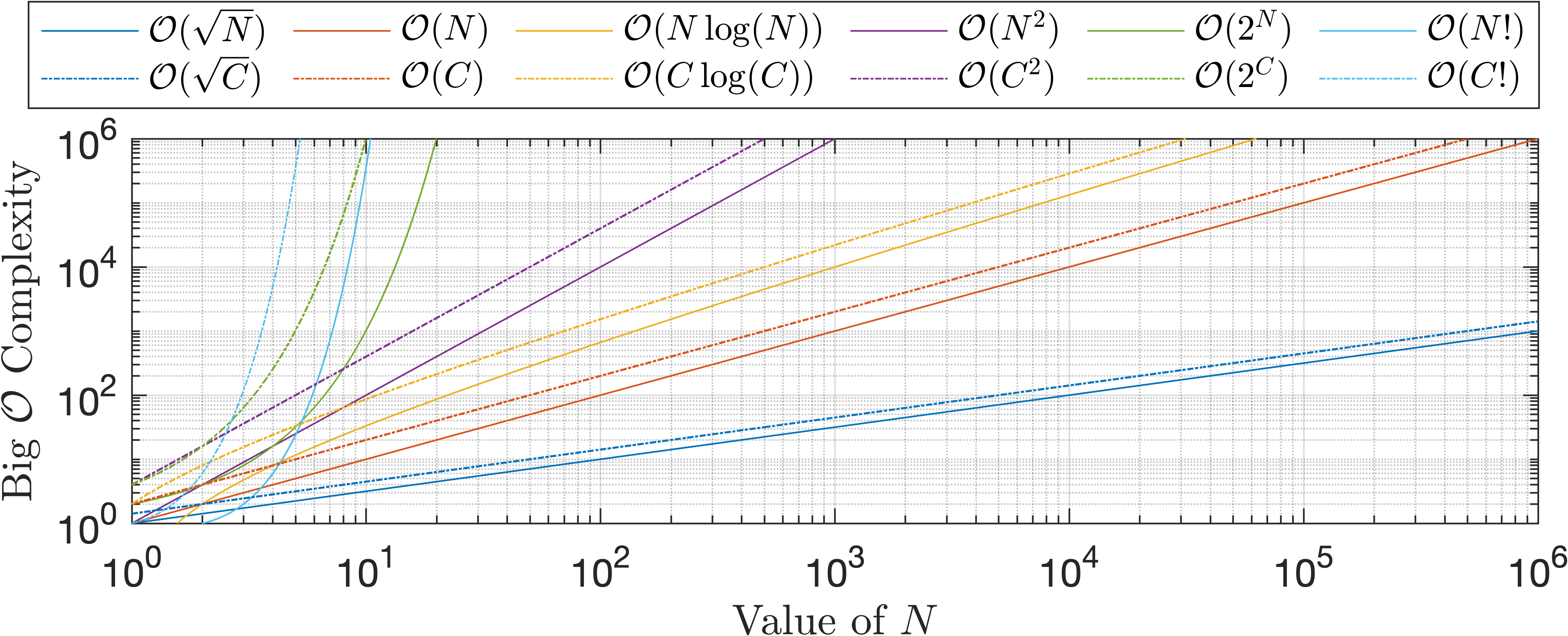}
  \caption{The computational and conversion complexity of problem classes on a logarithmic scale.}\label{fig:ComplexityCombined}
\end{figure}

Because the cost of getting data into an analog computing accelerator is high (Section~\ref{section:bottlenecks}), analog computing accelerators should target computationally-complex operations. 
If the operation an accelerator can accelerate has the same computational cost (for example $\mathcal{O}(N)$) as getting the data into and out of the accelerator (for example $\mathcal{O}(N)$) then such an accelerator will be constrained to produce a limited speedup over digital electronic hardware alone. 
Promising examples for acceleration are matrix-vector multiply accumulate operations $\mathcal{O}(N^2)$ and Ising problems $\mathcal{O}(2^{N})$.
Even when the computational complexity of a problem is larger than its input costs, it is still required that the problem is large enough to make the speedup worthwhile.
The compute-centric computational complexity metric does not capture the large data movement costs required to move data into computing accelerators.
The community of researchers investigating machine learning with new compute paradigms should instead adopt existing metrics for computational complexity that account for communications and data movement costs~\cite{KUSHILEVITZ1997331}.
Figure~\ref{fig:ComplexityCombined} shows a plot to illustrate the computational overhead introduced by the data conversions required to interface an analog computing accelerator with a digital electronic computer system.
Figure~\ref{fig:ComplexityCombined} assumes that the conversion complexity $C=2N$ as all $N$ data require a digital-to-analog conversion and then a subsequent analog-to-digital conversion.
In reality, the relationship between the computational complexity and the conversion complexity will depend upon the type of operations that are being accelerated and the implementation of the conversion and computing hardware.

\section{Bespoke Hardware Accelerators Require $10 \times$ Theoretical Improvement}\label{section:tentimes}

Designing and building a computing accelerator is time-consuming, expensive, and risky~\cite{graphcore_risk}.
Therefore, accelerators should provide at least $10 \times$ improvement of some metric that users care about for a large family of applications to be a commercial success~\cite{thiel2014zero}. 
In addition, the theoretical improvements produced by the accelerator must be large enough to absorb reductions in performance from the theoretical maximum caused by compromises in the design of the accelerator.  
The data conversion bottleneck in analog computing accelerators which Meech et al.~\cite{meech2023data} originally identified has recently been discussed in work on optical computing accelerators~\cite{anderson2023optical,mcmahon2023physics,zhong2023lightning,zhong2023demo}, thermodynamic~\cite{coles2023thermodynamic, aifer2023thermodynamic, duffield2023thermodynamic}, and neuromorphic computing accelerators~\cite{10272252}.
This article is the first to describe the data conversion bottleneck generally and its applicability for all analog computing accelerators.

\subsection{Theoretical Case Study: Analog Optical Fourier Transform and Convolution Accelerator}

Table~\ref{Table:Applications} and Figure~\ref{fig:speedUpBarChart} (Appendix~\ref{appendix:Amdahls}) show that an ideal optical accelerator in which Fourier transform and convolution operations cost zero time can only provide greater than $10 \times$ speedup for two of the benchmarked applications (pure convolutions and pure Fourier transforms). 
We found that the median end-to-end speedup achievable by an optical accelerator for 27 benchmark applications is $1.94 \times$, limited primarily by Amdahl's law (Appendix~\ref{appendix:Amdahls}).
This median speedup is small compared to the speedup achievable by other accelerators. 
The average speedup is $9.39 \times$, which is close to the $10 \times$ requirement to make the accelerator worthwhile (Section~\ref{section:tentimes}).
The high speedup values of $159.41 \times$ and $45.32 \times$ skew the average for convolutions and Fourier transforms. 
Our benchmarking study assumed zero cost for data movement, therefore our results are for the theoretical best case.

Popular accelerators in the literature report average speedups of $60 \times$ for convolutional neural networks on GPUs~\cite{latifi2016cnndroid}, $1.6 \times 10^{9}$ $\times$ for a quantum accelerator~\cite{arute2019quantum}, and $2076 \times$ fewer instructions executed compared to a Monte Carlo simulation for Laplace, an uncertainty quantification accelerator~\cite{tsoutsouras2021laplace}. 
These improvements are orders of magnitude larger than those theoretically possible with an optical accelerator. 
Therefore, developing an analog optical Fourier transform and convolution accelerator is not worthwhile unless we are targeting applications that consist solely of Fourier transforms and convolutions with less than $10$\,\% of execution time spent performing other operations; 
otherwise, by Amdahl's law, the acceleration is limited to less than 10-fold, the threshold below which it is not worth investing the time and capital in building an accelerator. 
A multiply-accumulate accelerator for neural network applications is a potentially more promising target for a commercial optical computing accelerator.
An optical physical computing accelerator implementation that accelerates the end-to-end inference latency of the LeNet deep neural network by $9.4 \times$ and $6.6 \times$ compared to Nvidia P4 and A100 graphics processing units respectively exists~\cite{zhong2023demo}.
The research article~\cite{zhong2023demo} which reports the inference speedup does not report an energy efficiency comparison.

\section{What Class of Computing Problems Suit Analog Computing Accelerators?}

\begin{figure}
  \centering
  \includegraphics[width=\textwidth]{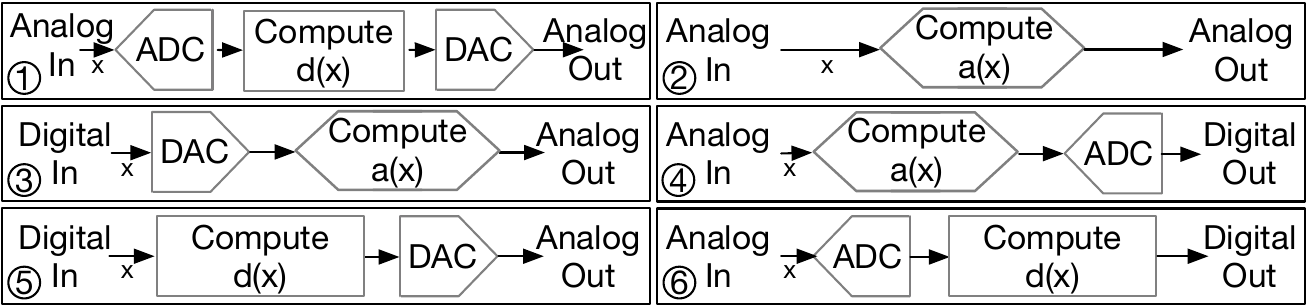}
  \caption{Architectures for computational problems with a variety of input and output data.
           Let $a(x)$ be an analog function, and $d(x)$ be a digital function both computed using the input data $x$.}\label{fig:sixTypes}
\end{figure}

Analog computing devices are best suited for performing computing problems with analog input and output data.
Figure~\ref{fig:sixTypes}~block~\ding{192} shows that two data conversions are required to use digital computing devices for a problem with analog input and output data.
Figure~\ref{fig:sixTypes}~block~\ding{193} shows that no data conversions are required to use analog computing devices to solve a computing problem with analog input and output data.
Therefore, using analog computing devices for computing problems with analog input and output data removes the data conversion overhead required to use digital computing devices.
For example, a well-known computing application with analog input and output is optically processing analog synthetic aperture radar images and then exposing analog camera film using the light output by the optical system~\cite{feitelson1988optical}.

When an application has analog input data and digital output data or vice versa we can choose to use analog or digital computing devices without incurring the penalty of an additional analog-to-digital or digital-to-analog conversion. 
Figure~\ref{fig:sixTypes} blocks~\ding{194},~\ding{195},~\ding{196},~and~\ding{197} show that we can choose to perform the computation before or after the conversion stage.
For this reason, researchers developing new novel analog computing devices should focus on accelerator architectures that follow the structure shown in Figure~\ref{fig:sixTypes}~blocks~\ding{193},~\ding{194},~and~\ding{195}. 
Sensor data processing applications that have the architecture shown in Figure~\ref{fig:sixTypes} block~\ding{195} are promising examples of applications where novel analog computing devices could have a high impact.
For example, an analog vision sensor data processing research implementation prevents the analog-to-digital conversion bottleneck by performing all processing on analog signals and converting the final output to digital~\cite{chen2023all}.
Waveform synthesis or control signal generation applications that have the architecture shown in Figure~\ref{fig:sixTypes} block~\ding{194} are promising examples of applications where novel analog computing devices could have a high impact.
Figure~\ref{fig:sixTypes} blocks~\ding{196}~and~\ding{197} show architectures suitable for applications of novel digital computing devices. 

\section*{Conclusion}

Modern computing tasks are constrained to having digital electronic input and output data.
Mass-produced electronic memory being the only off-the-shelf option for users, constrains the input data storage to be digital electronic signals stored in the memory. 
Support for plotting and data visualization software is only available for programming languages designed to run on off-the-shelf digital electronic hardware.
Therefore, any analog computing accelerator must perform an analog-to-digital conversion on its input data and a subsequent digital-to-analog conversion on its output data. 
The only alternative to this situation would be to develop an entire software and hardware stack to allow the analog computing devices to perform all the functions of the traditional digital electronic computer hardware. 
The traditional digital electronic computer architecture is better suited for the majority of applications than an application-specific analog computing accelerator and therefore substituting them would be unproductive. 
In a case study on an optical computing accelerator for Fourier transforms and convolutions we performed the first large-scale benchmarking of applications that rely on Fourier transform and convolution operations and found that the median end-to-end speedup achievable by an optical accelerator for 27 benchmark applications is $1.94 \times$, limited primarily by Amdahl's law (Appendix~\ref{appendix:Amdahls}).
This median speedup is small compared to the speedup achievable by other popular types of accelerators. 
The average speedup is $9.39 \times$, which is close to the $10 \times$ requirement to make the accelerator worthwhile (Section~\ref{section:tentimes}).
The high speedup values of $159.41 \times$ and $45.32 \times$ skew the average for convolutions and Fourier transforms. 
Our benchmarking study assumed zero cost for data movement, therefore our results are for the theoretical best case.
For optical accelerators to produce a worthwhile speedup we must overcome the data movement bottleneck. 
Once we have overcome the bottleneck, most applications will only be able to achieve a speedup of less than 10-fold. 
Our results show that building an analog optical Fourier transform and convolution accelerator is not worthwhile unless it will be applied to applications for which more than $90$\,\% of the execution time is Fourier transforms or convolutions. 
Even with faster light-modulating devices and camera detectors, the data movement bottleneck will continue to be a show-stopping problem for analog optical computing accelerators. 

\bibliographystyle{plain}
\bibliography{ms}

\appendix
\section{Analog Optical Fourier Transform and Convolution Accelerators}

Optical computing has been a popular research topic since the 1950s but there are still no commercially-available optical accelerators and no large-scale analysis of benchmark performance. 
Research implementations of optical computing accelerators and predictions of the performance of an application-specific integrated circuit implementation do however exist~\cite{zhong2023demo, zhong2023lightning, anderson2023optical}.
Despite this, there are no commercially available optical computing accelerators.
The physics of light lends itself to fast and efficient Fourier transform and convolution operations~\cite{Nicholas2017Reconfigurable,macfaden2017optical}:
Optical Fourier transform and convolution accelerators use diffraction, the interference of Huygens wavelets of light to perform Fourier transform operations~\cite{huygens1912treatise}. 
This is in contrast to digital electronic processors which break the high-level Fourier transform down into individual additions, multiplications, and other component operations, compute the results, and then recombine the results to calculate the Fourier transform~\cite{652905}.
Having the light perform the computation is faster and more efficient than using digital electronics if we do not consider the time required for data movement~\cite{macfaden2017optical}.

Despite these benefits, researchers in academic institutions and industry have struggled for 70 years to implement practically useful optical accelerators~\cite{caulfield2010future,Ambs2010,wilsonmultiply}. 
Startup companies repeatedly pivot to applying optical accelerators to new problems.
They do this because the optical accelerator does not provide a large enough improvement in a metric that users care about for the target application~\cite{cottle2020optical,wilsonmultiply,optalysys_history,optalysys_fully_homomorphic_encryption,meech2023computing}.
As of today, there is still no commercially available computer architecture that includes an optical accelerator, despite the growing popularity of optical interconnects~\cite{hotChipsLightMatter, hotChipsASOE}. 

\subsection{How Does an Analog Optical Fourier Transform and Convolution Accelerator Work?}

\begin{figure}
	\centering
	\includegraphics[width=\textwidth]{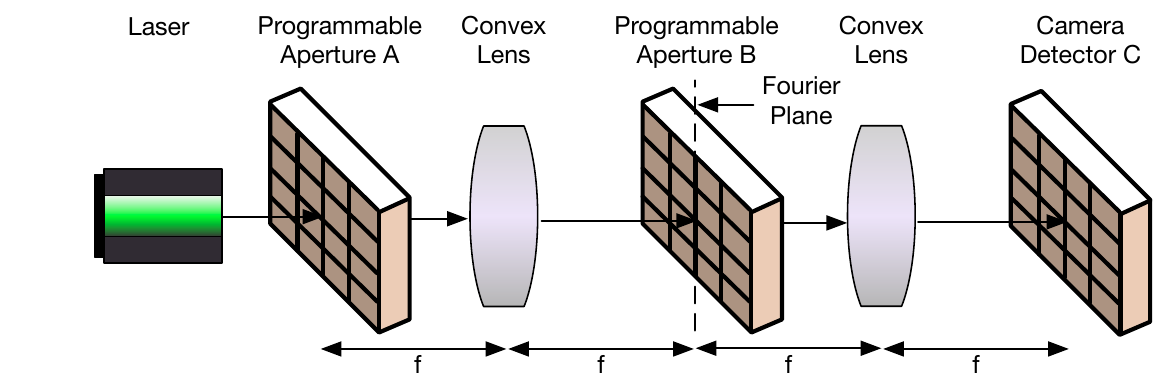}
	\caption{The 4$f$ setup for optical convolution where $A$ and $B$ are programmable apertures and $C$ is a camera detector.
			     Each optical component is spaced  a distance $f$ from the previous one where $f$ is the focal length of the convex lenses~\cite{Ambs2010}.}\label{fig:fourFsetup}
\end{figure}

Figure~\ref{fig:fourFsetup} shows the typical 4$f$ optical setup for Fourier transform and convolution operations. 
Let $\mathcal{F}$ be the Fourier transform operator and $\mathcal{F}^{-1}$ be the inverse Fourier transform operator.
Let $A$ and $B$ be two-dimensional arrays and $\mathcal{F}^{-1}{C}$ be the convolution of $A$ and $B$ where

\begin{equation}
	C = \mathcal{F}(A \circledast B) = \mathcal{F}(A) \cdot \mathcal{F}(B).
	\label{equationAcceleration}
\end{equation}

Equation~\ref{equationAcceleration} shows that an analog optical accelerator can perform the convolution operation by taking the Fourier transform of both input datasets, calculating their dot product, and finally, inverse Fourier transform the result. 
The optical setup cannot perform the final inverse Fourier transform step.
Instead, the digital electronic processor interfacing with the optical setup performs this step.
Figure~\ref{fig:fourFsetup} shows how the lenses in the setup Fourier transform the input data programmed into the aperture (spatial light modulator). 
The programmable aperture encodes information into the light at each of its pixels by manipulating the phase of the light between $0$ and $2 \pi$ according to the programmed digital value for that particular pixel.
An analog optical accelerator that uses a camera to transduce the output light pattern to electronic signals can only calculate the magnitude component of the right-hand side of Equation~\ref{equationAcceleration} and then the computer hardware must read the detector pixels and use a digital inverse Fourier transform to calculate the final result of Equation~\ref{equationAcceleration}. 
The light can only compute the Fourier transform when the condition (that $D \gg a$ and that $D \gg a^2 / \lambda$, 
where $D$ is the distance between the programmable aperture and the camera detector,
$a$ is the width of the programmable aperture, and $\lambda$ is the wavelength of the light~\cite{hecht2012optics}) for Fraunhofer diffraction is met~\cite{hecht2012optics}.

\subsection{Analog Optical Fourier Transform and Convolution Accelerator Computer Architecture}

\begin{figure*}[t]
	\centering
	\includegraphics[width=\textwidth]{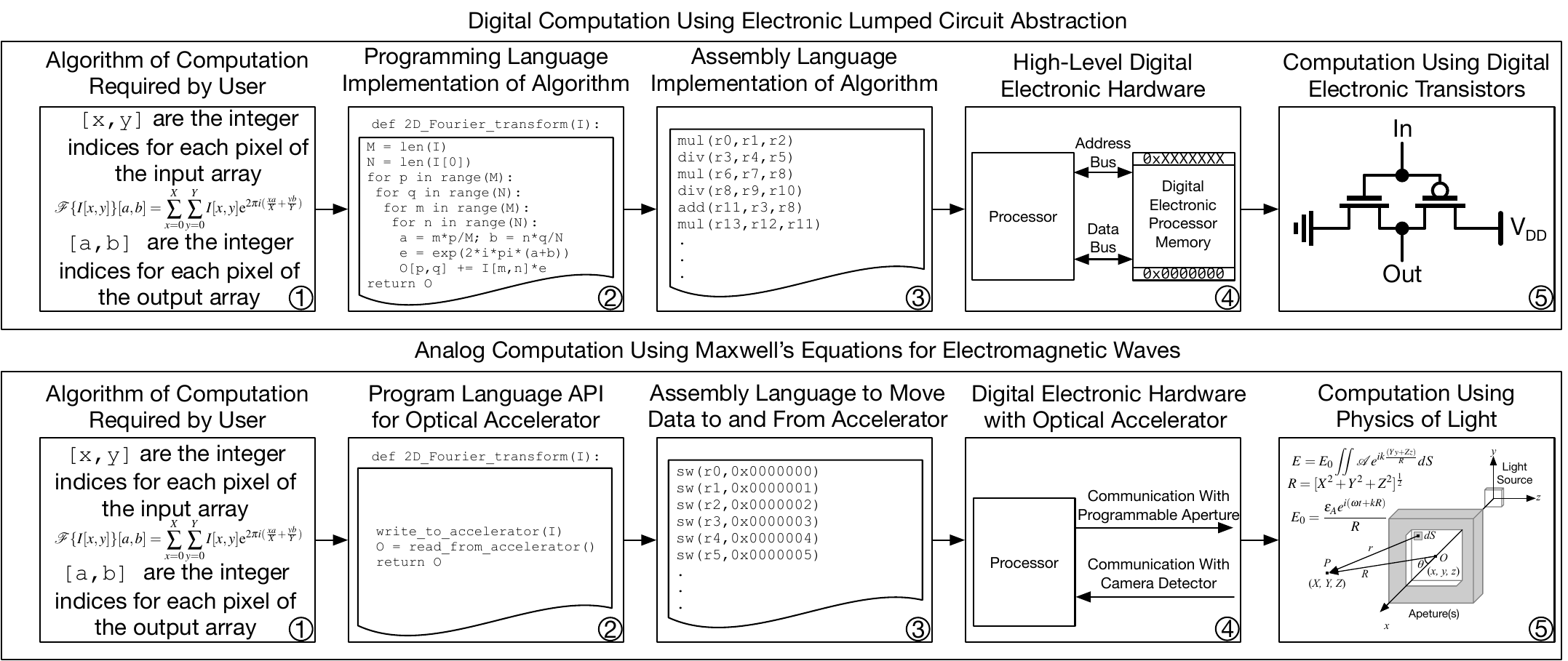}
	\caption{The steps required to perform a Fourier transform on data using an optical accelerator instead of a digital electronic processor diverge at the first abstraction level below the mathematical equation for the Fourier transform. 
	        The optical accelerator requires changes at every level of the software and hardware stack to use Maxwell's equations for electromagnetic waves to perform the Fourier transform.
	        This figure captures the idea that inspired 70 years of research into optical accelerators~\cite{hecht2012optics,baron1878analytical}.
			    The lumped circuit abstraction shown in row one confines the resistance, capacitance, and inductance of transistors within idealized circuit components.             
          This allows the designer to ignore the effects of electromagnetic waves. In contrast, row two directly uses the physics of electromagnetic waves to perform the computation.}\label{fig:bigPicture}
\end{figure*}

Figure~\ref{fig:bigPicture} shows the changes required at each abstraction layer of a software and hardware stack required to use the physics of light to accelerate a user-specified high-level computational problem (the Fourier transform).
A computer systems architect has to make changes at every abstraction level in the software and hardware stack to take advantage of the physics of light to perform computation. 
Required changes include a new software application programming interface to load data into the accelerator, processor architecture changes to allow store word and load word instructions to access the optical accelerator and digital electronic processor memory, and the close integration of optical hardware with digital electronic hardware that uses incompatible process technologies.
This is just as generations of engineers and scientists designed the modern digital electronic computer stack to realize the full potential of semiconductor transistors in digital electronic processors. 
Row one of Figure~\ref{fig:bigPicture} is the transition from the abstract idea of the Fourier transform through the abstraction layers to the digital electronic hardware that we wish to use to perform the computation.
Row two of Figure~\ref{fig:bigPicture} requires changes at every level of the software and hardware stack.
If we tried to use the physics of light to replace panel~\ding{196} of row one, the accelerator would not be able to use the Fourier transform properties of light and we would not see performance increases. 
Row two of Figure~\ref{fig:bigPicture} shows the missing implementations that have made such optical accelerators unnecessarily inefficient due to a lack of computer systems knowledge in the optical computing community and vice-versa.

The optical accelerator takes advantage of the physics of light to skip all of the component multiplication, division, and addition instructions shown in Figure~\ref{fig:bigPicture}, row one, block~\ding{194}. 
Instead, we load the data into the optical accelerator and the physics of light performs the Fourier transform computation in one analog step. 
The optical accelerator performs the transform using physics shown in Figure~\ref{fig:bigPicture}, row two, panel~\ding{196}. 
The optical field at point $P$ is the superposition of the optical field at each elemental area $dS$ of the total area, $S$, of the aperture. 
Every single point in the optical accelerator output contains information from every single point in the optical accelerator aperture input. 
Each point in the wavefront at the aperture produces Huygens wavelets and the optical field beyond the aperture is the superposition of all of the wavelets.
The similarities to the equation in block~\ding{192} of both rows of Figure~\ref{fig:bigPicture} are that the sum symbols use the value of each pixel in the input once per output pixel to compute the pixel-by-pixel result of the Fourier transform.
This skipping of steps provides an opportunity for the acceleration of Fourier transform and convolution operations provided that the cost of moving data into and out of the optical accelerator does not outweigh the speedup we gain by using the Fourier transform and convolution properties of light. 
Unfortunately, Section~\ref{section:bottlenecks} shows that the cost of moving data into and out of the optical accelerator will always be the bottleneck in analog optical Fourier transform accelerator designs. 
Appendix~\ref{appendix:benchmarks} shows that even the best-case speedup we can gain by using the analog Fourier transform and convolution properties of light is often small. 

\section{Optical Computing Accelerator Prototype Design and Construction}\label{appendix:physicalImplementation}

\begin{figure*}[t]
  \centering
  \begin{tabular}{c}
  \subfloat[The minimum architecture for an optical Fourier transform and convolution computing accelerator. 
        This architecture uses slow communication interfaces to move camera data into the processor and data from the processor into the spatial light modulator.
        These slow communications interfaces were designed for updating displays at 60\,\si{\hertz} and therefore bottleneck our accelerator.\label{fig:oldAccelerator}]{%
        \includegraphics[width=0.55\textwidth,keepaspectratio]{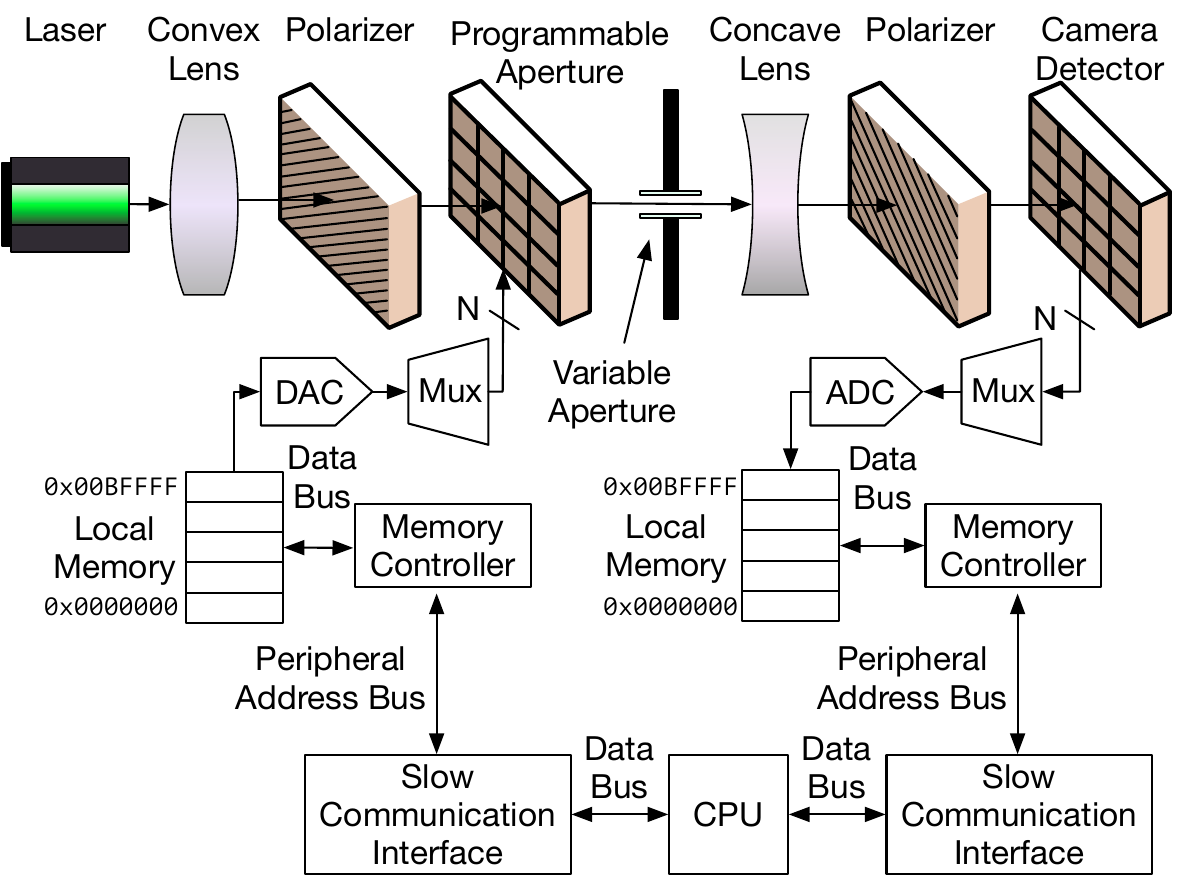}}
  \end{tabular}%
  \begin{tabular}{c}
  \subfloat[A side view of the prototype optical accelerator on an optical breadboard. The variable aperture is not programmable, only the spatial light modulator is programmable and controls the input data for the Fourier transform computation.\label{fig:side}]{%
        \includegraphics[width=0.40\textwidth,keepaspectratio]{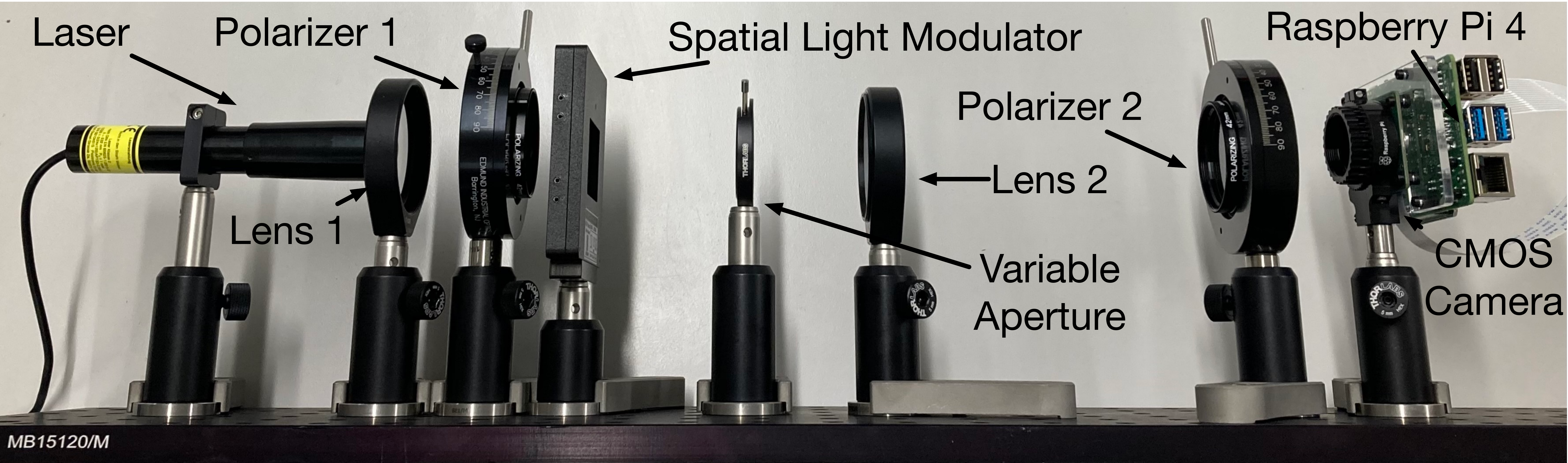}} \\
  \subfloat[A top view with components from left to right being the laser, polarizer, spatial light modulator, a lens to bring the far-field diffraction pattern closer to the laser, a second polarizer crossed with the first and the Raspberry Pi 4 mounted on the Raspberry Pi high-quality camera module.\label{fig:aerial}]{%
              \includegraphics[width=0.40\textwidth,keepaspectratio]{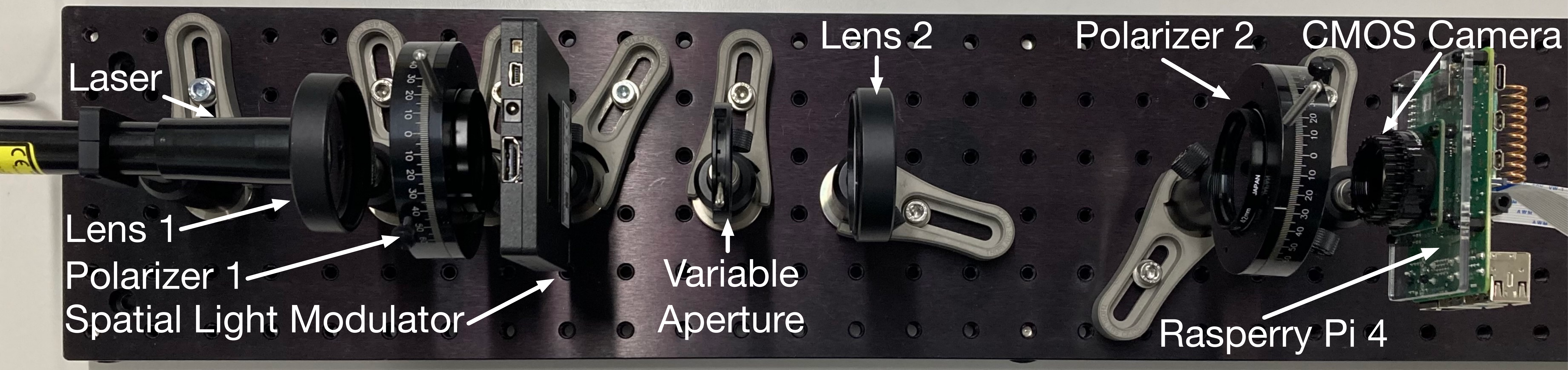}}   
  \end{tabular}
  \caption{The optical accelerator architecture diagram and the hardware prototype that we built to analyze the data-movement bottleneck.
       The Raspberry Pi 4 is an interface that we remotely connect to from a workstation computer using a secure shell and does not perform any computation other than programming the spatial light modulator and reading the camera.}\label{prototype}
  \end{figure*}
  
  Figure~\ref{fig:oldAccelerator} shows a block diagram of the typical interface between a digital electronic processor and an optical accelerator built using off-the-shelf optical hardware modules. 
  Typically these off-the-shelf optical hardware modules use a communication interface to allow a digital electronic processor to control the optical module as a peripheral input/output device. 
  Figure~\ref{fig:oldAccelerator} shows the local memory and digital-to-analog converter inside a spatial light modulator that allows an external digital electronic processor to program the light-modulating pixels over the communications interface. The camera provides a similar interface to allow the digital electronic processor to read values from the camera pixels.
  It uses an analog-to-digital converter to convert the analog signal from the camera detector pixels to a digital signal for the processor to read from the local device memory over the communication interface.  
  Spatial light modulators and digital micro-mirror devices are essentially a set of memory locations spatially arranged in large two-dimensional arrays. 
  Moving data from a processor into these memory locations and back costs time and energy. 
  This time and energy spent moving data outweigh the speed and efficiency benefits gained by using the properties of light to perform computation.
  
  Figures~\ref{fig:side}~and~\ref{fig:aerial} show our prototype implementation of a Fourier transform accelerator.
  We included the lenses, polarizers, and mechanical variable aperture to improve the resolution of the hardware prototype but they are not a fundamental requirement for performing Fourier transforms and convolutions using light.
  We conduct experiments to show the data-movement bottleneck using our hardware prototype.
  
  \subsection{Execution Time Experiment Methodology}
  
  We benchmark Python code to perform a $1024 \times 768$ pixel two-dimensional Fourier transform against the optical hardware setup performing the same calculation. 
  The hardware setup is an end-to-end system controlled by a Raspberry Pi 4 that runs Python scripts to activate the optical hardware. 
  For this reason, profiling the Python code quantifies the digital electronic processor, data movement, and analog optical accelerator computation time.
  
  \begin{figure}[t]
    \centering
    \includegraphics[width=\textwidth]{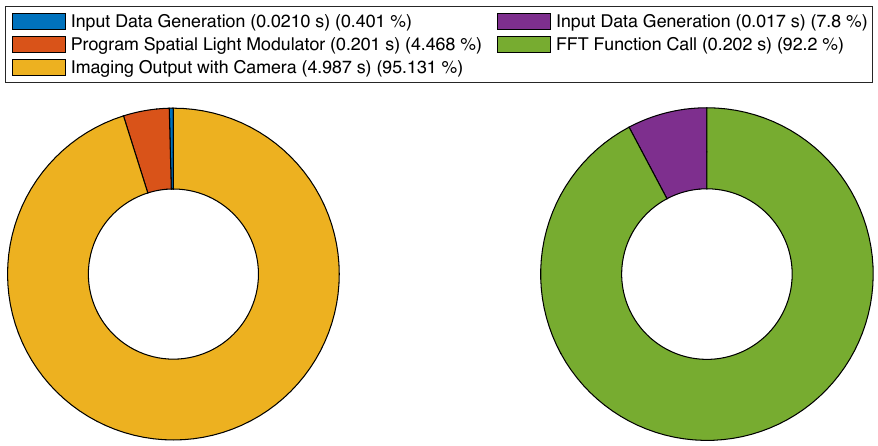}
    \caption{The hardware Fourier transform (left) is $23.8 \times$ slower than the NumPy software fast Fourier transform (right).
             The hardware Fourier transform takes negligible time compared to moving data into and out of the optical components. 
         The total time required to run the software and hardware Fourier transform is 0.219\,s and 5.209\,s respectively.}\label{fig:pie_charts}
  \end{figure} 
  
  \subsection{Execution Time Experiment Results}\label{section:executionTime}
  
  Figure~\ref{fig:pie_charts} shows that our off-the-shelf hardware prototype optical accelerator is $23.8 \times$ slower than a software fast Fourier transform of the same dimensions.
  We used the same Raspberry Pi 4 to benchmark the software fast Fourier transform and control the optical components (with no effort to optimize the code) alone to perform the Fourier transform. 
  As the Fourier transform computation happens at the speed of light, the only fixed computation that prevents infinite speedup (from Amdahl's law) is the time required to produce the input data, load it into the spatial light modulator, and then read out the output from the camera detector.
  The fast Fourier transform has the second-greatest theoretical speedup using an optical accelerator for all of the applications in Table~\ref{Table:Applications}.
  Therefore, none of the applications in Table~\ref{Table:Applications} will see a speedup when running on our prototype optical accelerator. 
  Figure~\ref{fig:pie_charts} shows that the majority of the computation time in the prototype optical accelerator is spent on data movement (programming the spatial light modulator and imaging the diffraction pattern using a camera).
  Boroumand et al.~\cite{boroumand2018google} state that 62.7\,\% of energy is spent on moving data in modern computing systems. 
  In our optical computing accelerator prototype 99.599\,\% of the time is spent moving data between the digital electronic processor and the analog optical accelerator.
  Cameras that can capture images significantly faster than the camera we used in our experiment exist~\cite{delbracio2021mobile}. 
  Nevertheless, the Fourier transform computation happens at the speed of light, so the data movement bottleneck will always dominate the computation time required by an optical Fourier transform and convolution computing accelerator.

\section{Convolution and Fourier Transform Benchmarking Case Study}\label{appendix:study}

\subsection{Convolution and Fourier Transform Application Benchmarking Methodology}

We profiled 27 benchmark applications (which we describe in Appendix~\ref{appendix:benchmarks}) to estimate the maximum theoretical speedup that an optical Fourier transform and convolution accelerator could provide for each application. 
We provide a short description of each benchmark that we profiled on a 2.8\,GHz Intel Core i7 CPU with 16\,GB of
2133\,MHz LPDDR3 RAM. 
All benchmarks are Python 3.8.9 code applications, not developed by the authors, which use well-optimized Python libraries, and are available online.
We used cProfile to profile each benchmark using Python 3.8.9 on MacOS Monterey Version 12.0.1.
We profiled each benchmark assuming that the time taken by functions with Fourier transform or convolution-related names was negligible.
We used the results to estimate the speedup gained by offloading the optical Fourier transform and convolution functions to an accelerator that completes the operation in negligible time. This assumption will provide results showing the best-case speedup for an optical Fourier transform and convolution accelerator. 

\subsection{Convolution and Fourier Transform Application Benchmarking Results: Amdahl's Law}\label{appendix:Amdahls}

We benchmarked the applications described in Appendix~\ref{appendix:benchmarks} using Python and cProfile and applied Amdahl's law to the results~\cite{Gustafson2011,amdahl1967validity}. 
We benchmarked each application one hundred times to take into account any variation.
Let $P$ be the degree of acceleration a computer system applies to an application, $f_\mathrm{fixed}$ be the portion of the program we cannot accelerate, and $f_\mathrm{accelerate}$ be the portion of the program that we can accelerate, then Amdahl's law states that the speedup $S$ we can achieve is 

\begin{equation}
	S = \frac{1}{f_\mathrm{fixed} + \frac{f_\mathrm{accelerate}}{P}}.
	\label{equation:amdahl}
\end{equation}

Using an optical accelerator to accelerate $f_\mathrm{accelerate}$  to the point that $\frac{f_\mathrm{accelerate}}{P} \ll f_\mathrm{fixed}$ produces 

\begin{equation}
	S \approx \frac{1}{f_\mathrm{fixed}}.
	\label{equation:amdahl2}
\end{equation}

\begin{figure*}
	\centering
	\includegraphics[width=\textwidth]{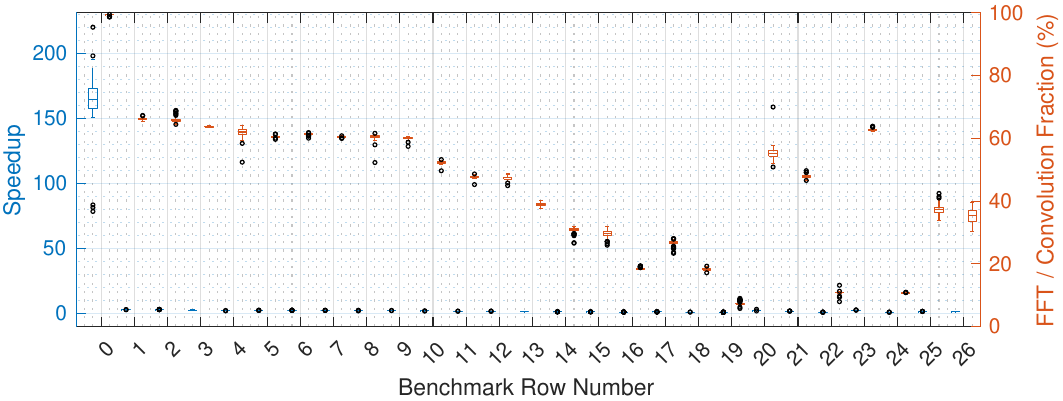}
	\caption{The potential end-to-end speedup for each application in Table~\ref{Table:Applications} according to Amdahl's law.
			 The speedups are small unless almost 100\,\% of end-to-end benchmark execution time is spent on Fourier transforms or convolutions.
			 The accelerator must speed up close to 100\% of the application code to produce a large end-to-end speedup.
			 All the box and whisker plots that show the run-to-run variation in the benchmark applications show small variation.
			 Box plot definitions: center line, median; box limits, upper and lower quartiles; whiskers, 1.5x interquartile range; points, outliers.}\label{fig:speedUpBarChart}
\end{figure*}

$S$ is the best case speedup we can achieve by accelerating the Fourier transform and convolution operations in a program. 
Figure~\ref{fig:speedUpBarChart} shows the potential speedup that we could get if we accelerated all Fourier transform and convolution operations in the benchmarks to the point where they were negligible. 
In practice, the speedup achieved by a real optical accelerator would be smaller because all optical accelerators require time for a digital electronic processor to write to the programmable aperture and read from the camera detector. 
Our benchmarking study has the unrealistic assumption that this writing and reading takes zero time. 
Table~\ref{Table:Applications} includes the names and descriptions of the benchmarks included in Figure~\ref{fig:speedUpBarChart}.

\begin{table*}
  \caption{The maximum end-to-end speedup achievable by an analog optical Fourier transform and convolution computing accelerator for a range of 27 different benchmark applications according to Amdahl's law.
       We ran each benchmark one hundred times and calculated the average for each column in the table.
       The average speedup is $9.39 \times$, close to the $10 \times$ requirement (Section~\ref{section:tentimes}).
       The result is heavily skewed by the high speedup values of $159.41 \times$ and $45.32 \times$ for convolutions and Fourier transforms. 
       The median speedup is $1.94 \times$ which is less than one-fifth of the $10 \times$ requirement.}\label{Table:Applications}
  \resizebox{\textwidth}{!}{%
  \begin{tabular}{cccccc}
  \toprule
  \cellcolor{a} Application & \begin{tabular}{@{}c@{}}\cellcolor{b} FFT/Conv \\\cellcolor{b} Time (s) \end{tabular} & \begin{tabular}{@{}c@{}}\cellcolor{a} Total \\ \cellcolor{a} Time (s) \end{tabular} & \begin{tabular}{@{}c@{}}\cellcolor{b} FFT/Conv \\ \cellcolor{b} Fraction (\%) \end{tabular} & \begin{tabular}{@{}c@{}}\cellcolor{a} End-to-End \\ \cellcolor{a} Speed Up ($\times$) \end{tabular} &\cellcolor{b} Lines \\\midrule 
  
  \rowcolor{b} Convolution~\cite{conv}                                                            & 0.158             & 0.159          & 99.37                  & 159.41              & 1     \\
  \rowcolor{a} Fourier Transform~\cite{fft}                                                       & 0.912             & 0.933          & 97.79                  & 45.32               & 1     \\
  \rowcolor{b} Wiener Filter~\cite{wienerFilter}                                                  & 1.164             & 1.724          & 67.51                  & 3.08                & 1     \\
  \rowcolor{a} Self-healing Airy beam~\cite{lightpipes}                                           & 51.718            & 81.778         & 63.24                  & 2.72                & 18    \\
  \rowcolor{b} Young's Experiment~\cite{lightpipes}                                               & 0.0671            & 0.109          & 61.70                  & 2.61                & 12    \\
  \rowcolor{a} From Poisson Spot to a Non-Diffractive Bessel Beam~\cite{lightpipes}               & 2.817             & 4.593          & 61.33                  & 2.59                & 20    \\
  \rowcolor{b} Generation of a Bessel Beam With a Lens and an Annular Slit~\cite{lightpipes}      & 3.146             & 5.173          & 60.82                  & 2.55                & 22    \\
  \rowcolor{a} Generation of a Bessel Beam With an Axicon~\cite{lightpipes}                       & 2.839             & 4.677          & 60.71                  & 2.55                & 18   \\ \rowcolor{b} Multi- holes and slits~\cite{lightpipes}                                           & 0.200             & 0.328          & 60.70                  & 2.55                & 21    \\
  \rowcolor{a} Diffraction From a Circular Aperture~\cite{lightpipes}                             & 2.193             & 3.615          & 60.65                  & 2.54                & 14    \\
  \rowcolor{b} Shack Hartmann Sensor~\cite{lightpipes}                                            & 2.142             & 4.051          & 52.88                  & 2.12                & 25    \\
  \rowcolor{a} Spot of Poisson~\cite{lightpipes}                                                  & 1.930             & 3.983          & 48.44                  & 1.94                & 12    \\
  \rowcolor{b} Fresnel Zone Plate~\cite{lightpipes}                                               & 0.665             & 1.405          & 47.34                  & 1.90                & 24    \\
  \rowcolor{a} Unstable Laser Resonator~\cite{lightpipes}                                         & 0.0645            & 0.163          & 39.43                  & 1.65                & 41    \\
  \rowcolor{b} Interference of a Doughnut Laser Beam: Collinear Beams~\cite{lightpipes}           & 0.0604            & 0.198          & 30.54                  & 1.44                & 16    \\
  \rowcolor{a} Michelson Interferometer~\cite{lightpipes}                                         & 0.0139            & 0.0472         & 29.45                  & 1.42                & 25    \\
  \rowcolor{b} Phase Recovery~\cite{lightpipes}                                                   & 0.296             & 1.580          & 18.75                  & 1.23                & 16    \\
  \rowcolor{a} \begin{tabular}{@{}c@{}}Transformation of a Fundamental Gauss Mode 															
  \\ into a Doughnut Mode With a Spiral Phase Plate~\cite{lightpipes} \end{tabular}               & 0.296             & 1.230          & 18.75                  & 1.23                & 13    \\
  \rowcolor{b} \begin{tabular}{@{}c@{}}Transformation of High Order 
  \\ Gauss Modes From Hermite to Laguerre~\cite{lightpipes} \end{tabular}                         & 0.0386            & 0.211          & 18.29                  & 1.22                & 42    \\
  \rowcolor{a} Interference of a Doughnut Laser Beam: Tilted Beams~\cite{lightpipes}              & 0.00506           & 0.0692         & 7.31                   & 1.08                & 15    \\
  \rowcolor{b} Double-Slit Experiment~\cite{prysym}                                               & 0.0519            & 0.0929         & 55.91                  & 2.27                & 12    \\
  \rowcolor{a} Your First Diffraction Model~\cite{prysym}                                         & 0.0787            & 0.164          & 47.80                  & 1.92                & 20    \\
  \rowcolor{b} Image Simulation~\cite{prysym}                                                     & 1.882             & 17.195         & 10.95                  & 1.12                & 45    \\
  \rowcolor{a} Convolutional Neural Network Inference~\cite{convCnn}                              & 0.263             & 0.416          & 63.17                  & 2.71                & 1     \\
  \rowcolor{b} Convolutional Neural Network Training~\cite{convCnn}                               & 8.428             & 78.936         & 10.68                  & 1.12                & 16    \\
  \rowcolor{a} Audio Resampling Transforms~\cite{audioResampling}                                 & 0.0513            & 0.135          & 37.94                  & 1.61                & 22    \\
  \rowcolor{b} Pre-Trained Model Wave2Vec2 Speech Recognition Inference~\cite{baevski2020wav2vec} & 0.179             & 0.519          & 34.53                  & 1.53                & 4     \\
  \bottomrule
  \end{tabular}}
  \end{table*}

\subsection{Convolution and Fourier Transform Benchmark Application Descriptions}\label{appendix:benchmarks}
\subsubsection*{Convolution (Application 0):} The SciPy implementation of convolution run over pre-generated $100 \times 100$ NumPy arrays. 
\subsubsection*{Fourier Transform (Application 1):} The NumPy fast Fourier transform implementation run over pre-generated $5000 \times 5000$ NumPy arrays. 
\subsubsection*{Wiener Filter (Application 2):} The SciPy implementation of the Wiener Filter run over a pre-generated $4000 \times 4000$ NumPy array. 
\subsubsection*{Self-healing Airy Beam (Application 3):} The LightPipes implementation of a self-healing Airy diffraction simulation.
Airy beams have applications including laser micromachining and particle and cell micro manipulation~\cite{fang20211d}.
\subsubsection*{Young's Experiment (Application 4):} The LightPipes implementation of a simulation of Young's double slit experiment.
In the experiment, a monochromatic plane wave illuminates two narrow slits which produces a diffraction pattern that illustrates the wave properties of light on a screen placed in the far field. 
The diffraction pattern is the Fourier transform of the slits function. 
It is possible to construct arbitrary far-field diffraction patterns by constructing the corresponding slit.
\subsubsection*{From Poisson Spot to a Non-Diffractive Bessel Beam (Application 5):} The LightPipes implementation of a simulation showing the proportionality of the width of a Bessel beam to the distance $z$ from the Huygens light point source. Bessel beams have applications in encryption, optical atom trapping, and optical tweezers~\cite{mcgloin2005bessel}.
\subsubsection*{Generation of a Bessel Beam with a Lens and an Annular Slit (Application 6):} The LightPipes implementation of a simulation of a Bessel beam.
Bessel beams have applications in encryption, optical trapping of atoms, and optical tweezers~\cite{mcgloin2005bessel}.
\subsubsection*{Generation of a Bessel Beam with an Axicon (Application 7):} Generating a Bessel beam with an annular slit is inefficient, most of the laser beam is unused.
This benchmark is the LightPipes implementation of generating a Bessel beam with an axicon lens that uses more of the total optical beam power than the annular slit method and is therefore, more efficient~\cite{cabrini2006axicon}.
\subsubsection*{Multi- Holes and Slits (Application 8):} The LightPipes implementation of a simulation of an extension of Young's experiment where multiple slits or holes are present. 
Changing the spacing and geometry of the holes would allow the user to create apertures that create arbitrary diffraction patterns and then simulate the resulting diffraction pattern. 
A multi-slit diffraction grating has applications as a spectrometer~\cite{kong2001infrared}.
\subsubsection*{Diffraction from a Circular Aperture (Application 9):} The LightPipes implementation of a simulation of an extension of Young's slit experiment where the aperture is circular instead of a slit. Diffraction through circular holes is used for simulating masks in epitaxy for semiconductors~\cite{hsu2012nanoepitaxy}.
\subsubsection*{Shack Hartmann Sensor (Application 10):} The LightPipes implementation of a Shack Hartmann sensor.
The Shack-Hartmann sensor is an array of lenses used to measure the phase distribution of a wavefront.
The US Air Force used them to improve the images of satellites taken from Earth~\cite{platt2001history}. 
\subsubsection*{Spot of Poisson (Application 11):} The LightPipes implementation of a simulation of a laser beam illuminating a disk. 
The result of the experiment is a bright spot of light directly behind the round disk. 
Poisson predicted the existence of the spot by applying Maxwell's equations, later Arago experimentally observed the spot. 
This was one of the first real-world demonstrations of the wave-like nature of light. 
The Arago spot has applications in the design of telescopes~\cite{cash2014ultra}.
\subsubsection*{Fresnel Zone Plate (Application 12):} The LightPipes implementation of the simulation of a Fresnel zone plate. 
The Fresnel zone plate acts as a focusing lens for a plane wave.
The Fresnel zone plate has applications in exoplanet detection~\cite{koechlin2005high}. 
\subsubsection*{Unstable Laser Resonator (Application 13):} The LightPipes implementation of the simulation of an unstable laser resonator. 
Unstable laser resonators build energy to create laser beams~\cite{1445615}.
\subsubsection*{Interference of a Doughnut Laser Beam Collinear Beams (Application 14):} The LightPipes doughnut laser with collinear beams interference simulation implementation.
\subsubsection*{Michelson Interferometer (Application 15):} The LightPipes implementation of a Michelson interferometer.
The Michelson interferometer has applications in spectrometers, measuring the diameter of stars, and detecting gravitational waves~\cite{michelson1887relative}.
\subsubsection*{Phase Recovery (Application 16):} The LightPipes implementation of the Gerchberg Saxton phase recovery algorithm. 
Phase recovery is the act of recovering electric field phase information that produces a diffraction pattern using only the light intensity of the diffraction pattern.
It iteratively performs forward and backward Fourier transforms and applies the constraints of the target intensity image until the algorithm converges to the phase of the electric field that produced the original image~\cite{gerhberg1972practical}. Phase recovery has applications in holography, electron microscopy, X-ray crystallography, and characterizing telescopes.
\subsubsection*{Transformation of a Fundamental Gauss Mode into a Doughnut Mode with a Spiral Phase Plate (Application 17):} 
The LightPipes implementation of a spiral phase plate simulation to produce a doughnut-shaped beam with applications in super-resolution microscopy, optical tweezers, and cell capture~\cite{watanabe2004generation}.
\subsubsection*{Transformation of High Order Gauss Modes From Hermite to Laguerre (Application 18):} The LightPipes implementation of a simulation that transforms Hermite Gauss into Laguerre Gauss laser modes using two cylindrical lenses.
Laguerre Gauss laser modes have applications in optical communication, micromanipulation, and quantum information~\cite{beijersbergen1993astigmatic}.
\subsubsection*{Interference of a Doughnut Laser Beam Tilted Beams (Application 19):} The LightPipes doughnut laser with tilted beams interference simulation implementation.
\subsubsection*{Double-Slit Experiment (Application 20):} The Prysm implementation of the simulation of Young's Experiment. 
The speedup value is similar to the LightPipes implementation.
\subsubsection*{Your First Diffraction Model (Application 21):} The Prysym implementation of diffraction through a circular aperture. 
The speedup value is similar to the LightPipes implementation.
\subsubsection*{Image Simulation (Application 22):} The Prysym implementation of an end-to-end image simulation of a Siemens' star including all optical and electrical noise. 
\subsubsection*{Convolutional Neural Network Inference (Application 23):} A PyTorch tutorial implementation of inference over a convolutional neural network for classifying images from the CIFAR10 dataset. 
We benchmarked the training and inference separately as they have significantly different potential potential for acceleration.
Convolutional neural networks have a wide range of applications~\cite{bhandare2016applications}. 
\subsubsection*{Convolutional Neural Network Training (Application 24):} A PyTorch tutorial implementation of training a convolutional neural network for classifying images from the CIFAR10 dataset. 
The speedup achieved for the training is less than half of the speedup achieved for the inference.
\subsubsection*{Audio Resampling Transforms (Application 25):} A PyTorch tutorial implementation of audio resampling using convolution. 
These transforms are used to resample audio before passing it through larger neural networks for training and inference. 
\subsubsection*{Pre-Trained Model Wave2Vec2 Speech Recognition Inference (Application 26):} A PyTorch implementation of speech recognition inference with the pre-trained Wave2Vec2 model.

\end{document}